\documentclass[twocolumn,showpacs,preprintnumbers,amsmath,amssymb,epsfig]{revtex4}

\usepackage{graphicx}
\usepackage{dcolumn}
\usepackage{bm}
\usepackage{epsfig}

\usepackage{float}
\usepackage{amsmath}
\usepackage{epsfig,floatflt}
\usepackage{subfigure}
\usepackage[usenames]{color}

\newcommand{\bea}{\begin{eqnarray}}
\newcommand{\eea}{\end{eqnarray}}

\def\bik{\mbox{\boldmath $k$}}

\begin{document}


\title{Axion as a cold dark matter candidate: low-mass case}

\author{Chan-Gyung Park${}^{1}$, Jai-chan Hwang${}^{2,3}$, and Hyerim Noh${}^{1}$}
\affiliation{${}^{1}$Korea Astronomy and Space Science Institute, Daejon 305-348, Republic of Korea \\
                 ${}^{2}$Department of Astronomy and Atmospheric Sciences,
                 Kyungpook National University, Daegu 702-701, Republic of Korea \\
                 ${}^{3}$Korea Institute for Advanced Study, Seoul 130-722, Republic of Korea}
\date{\today}


\begin{abstract}
Axion as a coherently oscillating scalar field is known to behave
as a cold dark matter in all cosmologically relevant scales. For
conventional axion mass with $10^{-5}~\textrm{eV}$, the axion reveals a
characteristic damping behavior in the evolution of density perturbations
on scales smaller than the solar system size. The damping scale is
inversely proportional to the square-root of the axion mass.
We show that the axion mass smaller than $10^{-24}~\textrm{eV}$ induces
a significant damping in the baryonic density power spectrum in
cosmologically relevant scales, thus deviating from the cold dark matter
in the scale smaller than the axion Jeans scale.
With such a small mass, however, our basic assumption about the coherently
oscillating scalar field is broken in the early universe.
This problem is shared by other dark matter models
based on the Bose-Einstein condensate and the ultra-light scalar field.
We introduce a simple model to avoid this problem by introducing
evolving axion mass in the early universe, and present observational effects of
present-day low-mass axion on the baryon density power spectrum,
the cosmic microwave background radiation (CMB) temperature power spectrum,
and the growth rate of baryon density perturbation.
In our low-mass axion model we have a characteristic
small-scale cutoff in the baryon density power spectrum
below the axion Jeans scale.
The small-scale deviations from the cold dark matter model
in both matter and CMB power spectra clearly differ from the ones expected
in the cold dark matter model mixed with the massive neutrinos
as a hot dark matter component.
\end{abstract}

\maketitle

\section{Introduction}

Observations of type Ia supernovae, cosmic microwave background (CMB)
radiation anisotropy, and large-scale structure of galaxies have revealed that
our present universe is dominated by two dark components \cite{Recent-Obs}.
Based on the Friedmann world model, the energy content
of the present universe is occupied by 73\% of dark energy and 23\% of
dark matter, whose nature is the fundamental mystery of the present day
theoretical physics and cosmology
(see Ref.\ \cite{Komatsu-etal-2011} for recent cosmological interpretation
of astronomical observations).

The dark matter is often modeled as pressureless, non-relativistic particles,
which is known as cold dark matter (CDM).
Although the CDM model has been greatly successful in explaining many
cosmological observations, there remain certain conflicts at the galactic
scales. First, the CDM-based simulations of galactic halos show density
profiles with a central cusp at kpc scale, while observations indicate
constant density cores \cite{vanEymeren-etal-2009}.
Secondly, the CDM model predicts overpopulation of low mass halos within
a galaxy or a galaxy group, which is an order of magnitude larger than
the present observations \cite{Klypin-etal-1999}.

These CDM problems at galactic scales may be overcome by introducing
a suppression of small-scale power in the density perturbation.
There have been efforts to explain the CDM conflicts using the
dark matter models with the Bose-Einstein condensate (BEC) and
the ultra-light scalar field.

In the BEC dark matter model the bosonic particles occupy the lowest
quantum state of the external potential and can be described as a
coherent matter wave of macroscopic size due to Bose enhancement
\cite{Sin-1994,Lee-1996-2010,BEC-refs,Rodriguez-Montoya-2010,
Kain-2012,Chavanis-2012}. 
Along this line there is fuzzy cold dark matter model proposed
by Hu et al.\ \cite{Hu-2000} which is a case of free particles
ignoring the self-interaction terms in the BEC based on the generalized
dark matter approach \cite{Hu-1998}.
In this model the coherent wave of ultra-light dark matter with
$m \sim 10^{-22}~\textrm{eV}$ can suppress the kpc-scale cusp in the dark matter
halos and reduce the abundance of low mass halos.
Recent investigations on the growth of perturbation in the universe
with BEC dark matter can be found in Refs.\ \cite{Kain-2012,Chavanis-2012}.
The BEC dark matter has a close connection to the axion dark matter model.
Recently, Sikivie and Yang \cite{Sikivie-2009} showed that
the cold dark matter axions form a BEC through their self-interactions
and gravitational interactions (see also Ref.\ \cite{Erken-2012}).

Another possibility is the ultra-light scalar field dark matter model
\cite{Peebles-1999,Matos-1999,Sahni-2000,Matos-2000-2001,Arbey-2001,
Matos-2009,Marsh-2010,Marsh-2011,Suarez-2011,Magana-2012}
(see Ref.\ \cite{Magana-2012-review} for a review).
Based on a specific scalar field potential with appropriate initial conditions
the evolution of scalar field mimics the property of the CDM fluid.
For example, authors of Refs.\ \cite{Sahni-2000,Matos-2000-2001}
proposed $\cosh$-like potential $V=V_0 (\cosh \lambda\phi-1)$
as a good candidate of dark matter (see also Ref.\ \cite{Matos-2009}).
Authors of Refs.\ \cite{Suarez-2011,Magana-2012} investigated the structure
formation in the scalar-field dark matter model with
$V=\frac{1}{2}m^2\phi^2+\frac{1}{4}\lambda\phi^4$ by using the fluid and
the field approaches. Authors of Refs.\ \cite{Marsh-2010,Marsh-2011}
studied the effect of ultra-light scalar field with $V=\frac{1}{2}m^2\phi^2$
on the growth of structure of the universe,
in a situation where only a substantial fraction of ultra-light scalar field
dark matter constitutes the total dark matter including CDM.

It is well known that the axion as a coherently oscillating scalar field
behaves as CDM in cosmologically relevant scales \cite{Axion-refs}.
Khlopov et al.\ \cite{Khlopov-etal-1985} investigated the growth of
perturbations due to gravitational instability in a universe dominated
by a scalar axion field.
The quantum description of cosmological axion perturbations was given
by Nambu and Sasaki \cite{Nambu-1990}, while the full relativistic description
based on the linear perturbation theory was presented in Refs.\
\cite{Ratra-1991,Hwang-1997,Hwang-2009}. For a typical mass with
$10^{-5}~\textrm{eV}$ the behavior of the axion is equivalent to that of
CDM on the cosmological scales including the superhorizon scale \cite{Hwang-2009}.

In this work, we consider an axion dark matter model
with extremely low mass ($m \le 10^{-22}~\textrm{eV}$) and investigate
the effect of ultra-light axion on the baryon matter density and the CMB
anisotropy power spectra and the perturbation growth.
Our calculations are based on the full relativistic linear cosmological
perturbation theory. Note that some of previous studies on BEC dark matter
models relies on Newtonian cosmology and the relativistic description
was given only in a limited fashion.
Due to the common property of coherent oscillations,
our description for the low-mass axion is closely related with that from
the BEC or ultra-light scalar field dark matter models.

The structure of this paper is as follows.
Section \ref{sec:basic_equations} reviews the basic equations
used to describe background and perturbation evolutions based on the
relativistic linear perturbation theory, dealing with the multi-component
fluids together with the minimally coupled scalar field.
In Sec. \ref{sec:axion} the basic properties of the axion dark matter are
briefly described. A complete set of perturbation equations for the axion,
baryon, and radiation fluids is presented, and initial conditions for
the perturbation variables in the early radiation era are derived for
two asymptotic limits. Our primary results are summarized in Sec.\
\ref{sec:results}, including baryon matter density and CMB power spectra
and the growth of baryon density perturbation for various values of
extremely light axion mass. We also compare the cosmological effects
of light axion with the CDM model mixed with the massive neutrinos
as a hot dark matter component. We discuss our results in Sec.\
\ref{sec:discussion}. Throughout this paper, we set $c\equiv 1 \equiv \hbar$.

\section{Basic equations}
\label{sec:basic_equations}

We consider scalar-type perturbations with the metric \cite{Bardeen-1988}
\bea
   & & ds^2 = -(1+2\alpha)dt^2-2a\beta_{,\alpha}dtdx^\alpha
   \nonumber \\
   & & \qquad
         + a^2 \left[(1+2\varphi)g_{\alpha\beta}^{(3)}
         + 2 \gamma_{,\alpha|\beta} \right] dx^\alpha dx^\beta,
\eea
and introduce
\bea
   \chi\equiv a(\beta+a\dot\gamma), \quad
       \kappa \equiv 3H\alpha -3\dot\varphi-{\Delta \over a^2} \chi.
\eea
Here $a(t)$ is the cosmic scale factor, $\alpha$, $\beta$,
$\varphi$, $\gamma$ are metric perturbations, a dot and a comma indicate
the time and spatial derivatives, respectively, and a vertical bar
represents the covariant derivative based on the comoving three-space metric
$g_{\alpha\beta}^{(3)}$.
The energy-momentum tensor is
\bea
   & & T^0_0 = - \mu - \delta \mu, \quad
       T^0_\alpha = - {1 \over k} \left( \mu + p \right) v_{,\alpha},
   \nonumber \\
   & & T^\alpha_\beta = \left( p + \delta p \right) \delta^\alpha_\beta,
\eea where we ignore the anisotropic stress;
$k$ is the wave number with $\Delta \equiv - k^2$ in Fourier space.

The background evolution is described by
\begin{eqnarray}
   && H^2
       = \frac{8\pi G}{3} \mu -\frac{K}{a^2} + \frac{\Lambda}{3}, \nonumber \\
   && \dot\mu + 3H(\mu+p)=0,
\end{eqnarray}
where $H \equiv \dot a/a$ is the Hubble parameter, $K$ is the sign of spatial curvature, and 
$\Lambda$ is the cosmological constant.
In the multiple component case we have
\begin{eqnarray}
  && \mu=\sum_j{\mu_j}, \quad p=\sum_j{p_j},
\end{eqnarray}
and
\bea
  && \dot\mu_i + 3H(\mu_i+p_i)=0,
\eea
where we ignore direct interactions among fluids.
For a minimally coupled scalar field we have the equation of motion
\bea
  & & \ddot\phi+3H\dot\phi+V_{,\phi}=0,
\eea
and the fluid quantities
\bea
  & & \mu_\phi=\frac{1}{2}\dot\phi^2+V,\quad
       p_\phi  =\frac{1}{2}\dot\phi^2-V.
\eea
We need the following equations for the perturbations \cite{Bardeen-1988,Hwang-1991}
\bea
   & & \dot\kappa+2H\kappa-4\pi G(\delta\mu+3\delta p)
         +\left(3\dot{H}-\frac{k^2}{a^2}\right)\alpha = 0,
   \label{eq:kappa_dot} \\
   & & \delta\dot\mu_i+3H(\delta\mu_i+\delta p_i)
       -(\mu_i+p_i)\left(\kappa-3H\alpha-\frac{k}{a}v_i\right)
   \nonumber \\
   & & \qquad
         =0,
   \label{eq:mu_dot} \\
   & & \frac{1}{a^4} \left[a^4 (\mu_i+p_i)v_i \right]^{\displaystyle\cdot}
        = \frac{k}{a} \left[ \left( \mu_i + p_i \right) \alpha
        + \delta p_i \right],
   \label{eq:v_dot}
\eea
where
\bea
   & & \delta\mu=\sum_j \delta\mu_j, \quad \delta p=\sum_j \delta p_j,
   \nonumber \\
   & & (\mu+p)v=\sum_j (\mu_j+p_j) v_j.
\eea
In the presence of a scalar field we have the equation of motion \cite{Hwang-1991}
\bea
   & & \delta\ddot\phi+3H\delta\dot\phi +\frac{k^2}{a^2}\delta\phi
         +V_{,\phi\phi}\delta\phi
   \nonumber \\
   & & \qquad
         =2\ddot\phi\alpha
         +\dot\phi \left(\dot\alpha+6H\alpha+\frac{k^2}{a^2}\chi
         -3\dot\varphi\right),
    \label{eq:field_eom}
\eea
and the fluid quantities
\bea
   & &  (\mu_\phi+p_\phi)v_\phi=\frac{k}{a}\dot\phi\delta\phi, \quad
          \delta\mu_\phi=\dot\phi\delta\dot\phi-\dot\phi^2\alpha
                       +V_{,\phi}\delta\phi,
   \nonumber \\
   & & \delta p_\phi=\dot\phi\delta\dot\phi-\dot\phi^2\alpha-V_{,\phi}\delta\phi,
   \label{eq:field_fluid}
\eea
with vanishing anisotropic stress.

\section{Axion}
\label{sec:axion}

We consider the axion in the presence of baryonic dust and radiation (photons and neutrinos).
The axion is considered as a minimally coupled scalar field with a potential
$V=\frac{1}{2}m^2\phi^2$, where $m$ is the axion mass.
We strictly ignore $H/m$ higher order term in our analysis.
At the current epoch it is
\begin{equation}
   \frac{H_0}{m} = 2.133\times 10^{-28} h
               \left(\frac{m}{10^{-5}~\textrm{eV}}\right)^{-1},
\end{equation}
where $H_0 \equiv 100 h~\textrm{km}\textrm{s}^{-1}\textrm{Mpc}^{-1}$ is
the Hubble expansion rate at present.

In the axion model the equation of motion leads to the background solution
as \cite{Ratra-1991}
\begin{equation}
   \phi(t)=a^{-3/2} [\phi_{+0}\sin(mt)+\phi_{-0}\cos(mt)],
\end{equation}
where $\phi_{+0}$ and $\phi_{-0}$ are constant coefficients.
We use the fluid formulation of axion. 
We have
\begin{equation}
   \mu_a=\frac{m^2}{2a^3} (\phi_{+0}^2+\phi_{-0}^2), \quad p_a=0,
\end{equation}
thus, the axion behaves as a zero-pressure fluid; for the fluid quantities
the time-averaging has been applied for highly oscillating scalar field.
For baryon, radiation (as a fluid), and axion, we have
\begin{equation}
   \mu=\mu_b + \mu_r + \mu_a, \quad p=\frac{1}{3}\mu_r.
\end{equation}

To perturbed order, with an ansatz \cite{Ratra-1991}
\begin{equation}
   \delta\phi(\bik,t)=\delta\phi_{+}(\bik,t)\sin(mt)
                     +\delta\phi_{-}(\bik,t)\cos(mt),
\end{equation}
the fluid quantities in Eq.\ (\ref{eq:field_fluid}) give \cite{Hwang-2009}
\bea
  & & \mskip-36mu \delta\mu_a =
        a^{-3/2} m [ m(\phi_{+0}\delta\phi_{+}+\phi_{-0}\delta\phi_{-}),
  \nonumber \\
  & & \mskip-36mu \qquad
        +\frac{1}{2}(\phi_{+0}\delta\dot\phi_{-}
        -\phi_{-0}\delta\dot\phi_+)]-\mu_a\alpha,
  \nonumber \\
  & & \mskip-36mu \delta p_a = \frac{1}{2} a^{-3/2} m
        (\phi_{+0}\delta\dot\phi_{-}-\phi_{-0}\delta\dot\phi_{+})
        -\mu_a \alpha,
  \nonumber \\
  & & \mskip-36mu {a \over k} (\mu_a+p_a)v_a=\frac{1}{2}a^{-3/2}m
        (\phi_{+0}\delta\phi_{-}-\phi_{-0}\delta\phi_{+}),
\label{eq:field_fluid_avg}
\eea
where for the fluid quantities we have taken the time average.
Up to this point the perturbed quantities are spatially gauge invariant and
the temporal gauge condition has not been taken yet \cite{Bardeen-1988}.

\subsection{Axion-comoving gauge}

As the temporal gauge (hypersurface or slicing) condition we
take the axion-comoving gauge \cite{Hwang-2009}
\begin{equation}
   \left( \mu_a + p_a \right) v_a \equiv 0.
\end{equation}
Equation (\ref{eq:field_fluid_avg}) gives
\begin{equation}
   \frac{\delta\phi_+}{\phi_{+0}}=\frac{\delta\phi_{-}}{\phi_{-0}}
\end{equation}
and
\begin{equation}
   \frac{\delta\mu_a}{\mu_a}
     =2 a^{3/2} \frac{\delta\phi_+}{\phi_{+0}}-\alpha,
   \quad
   \frac{\delta p_a}{\mu_a}=-\alpha.
\label{eq:field_fluid_v}
\end{equation}
From Eqs.\ (\ref{eq:kappa_dot})
and (\ref{eq:mu_dot}) for the axion, we have
\bea
    & & \dot\kappa+2H\kappa
          =4\pi G(\mu_b\delta_b+2\mu_r\delta_r+\mu_a\delta_a)
    \nonumber \\
    & & \qquad
          + \left[12\pi G \left(\mu_b+\frac{4}{3}\mu_r \right)
          +\frac{k^2-3K}{a^2} \right]\alpha,
    \label{eq:kappa_dot2} \\
    & & \dot\delta_a=\kappa,
\label{eq:delta_dot2}
\eea
where $\delta_i\equiv \delta\mu_i/\mu_i$.
For the baryon and radiation, under the fluid formulation,
Eqs.\ (\ref{eq:mu_dot}) and (\ref{eq:v_dot}) give
\begin{eqnarray}
   && \dot\delta_b = \kappa-3H\alpha-\frac{k}{a}v_b,
   \\
   && \dot{v}_b+Hv_b=\frac{k}{a}\alpha,
   \\
   && \dot\delta_r=\frac{4}{3}\left(\kappa-3H\alpha-\frac{k}{a}v_r\right),
   \\
   && \dot{v}_r = \frac{k}{a}\left(\alpha+\frac{1}{4}\delta_r\right).
   \label{eq:v_dot2}
\end{eqnarray}
For a zero-pressure fluid (like a CDM) we have $\alpha=-\delta p/\mu=0$.
Now, in the case of axion, $\alpha$ can be determined in terms of $\delta_a$
using Eq.\ (\ref{eq:field_eom}). From Eq.\ (\ref{eq:field_fluid}) we have
\begin{equation}
   \delta_a=2 a^{3/2}\frac{\delta\phi_+}{\phi_{+0}}-\alpha, \quad
   \alpha=-\frac{a^{3/2}}{2m^2}\frac{k^2}{a^2}\frac{\delta\phi_+}{\phi_{+0}}.
\end{equation}
From these two relations, we have \cite{Hwang-2009}
\begin{equation}
   \alpha=-\frac{k^2}{4m^2a^2}
           \frac{1}{1+\frac{k^2}{4m^2a^2}}\delta_a.
\label{eq:alpha}
\end{equation}
Equations (\ref{eq:kappa_dot2})--(\ref{eq:v_dot2}) and
(\ref{eq:alpha}) constitutes the complete set for axion, baryon, and radiation.
In a realistic situation we have to use the Boltzmann equations
for radiation (photons and neutrinos) or the tight-coupling approximation
for photon-baryon fluid \cite{Ma-Bertschinger-1995,Hwang-2001}. We note that the above equations are valid
in the presence of $K$ and $\Lambda$ in the background.

From Eqs.\ (\ref{eq:kappa_dot2}), (\ref{eq:delta_dot2})
and (\ref{eq:alpha}) we have
\bea
   & & \ddot \delta_a + 2 H \dot\delta_a
         - 4 \pi G \left( \mu_a \delta_a
         + \mu_b \delta_b + 2 \mu_r \delta_r \right)
   \nonumber \\
   & & \qquad
         + \left[ 12 \pi G \left( \mu_b + {4 \over 3} \mu_r \right)
         + \frac{k^2-3K}{a^2} \right]
   \nonumber \\
   & & \qquad
         \times
         \frac{k^2}{4m^2a^2}
         \frac{1}{1+\frac{k^2}{4m^2a^2}}\delta_a
         = 0.
\eea
The equation is valid at {\it all} scales.
Ignoring $K$, and for axion only, we obtain the axion Jeans scale,
\begin{equation}
   \lambda_J \equiv \frac{2\pi a}{k_J}
      = \frac{2\pi}{\sqrt{2\pi G\mu_a
                         +\sqrt{4\pi^2 G^2 \mu_a^2+16\pi G\mu_a m^2}}}.
\label{eq:Jeans_scale_general}
\end{equation}
In the limit of $k^2 / (4m^2a^2) \ll 1$, with present
$\Omega_{a} \equiv 8 \pi G \mu_a / (3H^2) \simeq 0.27$,
the current Jeans scale becomes
\cite{Khlopov-etal-1985,Nambu-1990,Sikivie-2009,Hwang-2009}
\begin{equation}
\begin{split}
   \lambda_J
       =\left(\frac{\pi^3}{G\mu_{a0} m^2}\right)^{1/4}
       = 50 h^{-1/2} \left(\frac{m}{10^{-5}~\textrm{eV}}\right)^{-1/2}
            ~\textrm{AU},
\end{split}
\label{eq:Jeans_scale}
\end{equation}
which is about the solar system size for $m\sim 10^{-5}~\textrm{eV}$.
For extremely low mass axion, we have
\begin{equation}
   \lambda_J = 2.4 h^{-1/2} \left(\frac{m}{10^{-25}~\textrm{eV}}\right)^{-1/2}
            ~\textrm{Mpc},
\label{eq:Jeans_scale2}
\end{equation}
which is a cosmologically significant scale.
On scales larger than the axion Jeans scale the axion behaves as the CDM.
In the following we are interested in the observable consequences of such
a small mass axion as a dark matter candidate.

\subsection{Initial conditions}

In order to have initial conditions for perturbation variables in the early
radiation dominated era,
we consider the baryon, radiation, and axion in a flat background.
In the radiation dominated era ($a\propto t^{1/2}$),
Eqs.\ (\ref{eq:kappa_dot2})-(\ref{eq:v_dot2}) give
\begin{eqnarray}
 && \ddot\delta_a+\frac{1}{t}\dot\delta_a
     =\frac{3}{4t^2}\delta_r+\left(\frac{3}{2t^2}+\frac{k^2}{a^2}\right)\alpha,
     \nonumber \\
 && \ddot\delta_r+\frac{1}{2t}\dot\delta_r
     +\left(-\frac{1}{t^2}+\frac{k^2}{3a^2}\right)\delta_r
     =-\frac{2}{3t}\dot\delta_a-\frac{2}{t}\dot\alpha+\frac{3}{t^2}\alpha,
     \nonumber \\
 && \ddot\delta_b+\frac{1}{t}\dot\delta_b
     =\frac{3}{4t^2}\delta_r-\frac{3}{2t}\dot\alpha+\frac{3}{2t^2}\alpha.
\end{eqnarray}
We consider asymptotic solutions in the large-scale limit, thus
$k/(aH) \ll 1$.

(I) For $k^2/(m^2 a^2) 
\ll 1$, we have $\alpha=0$; this is
the same as CDM. For $\delta_a \propto t^n$ we have
$(n-1)(2n-1)n(n+1)=0$, thus $n=1$, $\frac{1}{2}$, $0$, $-1$, and
the growing solution is
\begin{equation}
   \delta_a = \delta_b = \frac{3}{4} \delta_r \propto t.
\label{eq:ic1}
\end{equation}
This is the same as the adiabatic solution in the case of CDM.

(II) For $k^2/(m^2 a^2) 
\gg 1$, we have $\alpha=-\delta$.
For $\delta_a \propto t^n$, we have $(n-1)(n-\frac{1}{2})(n^2+n+\frac{3}{2})=0$,
and the growing mode is
\begin{equation}
   \delta_a = \frac{2}{5}\delta_b=\frac{3}{10}\delta_r \propto t.
\label{eq:ic2}
\end{equation}
The condition $k/(ma) \gg 1$ together with $k/(aH) \ll 1$ demands
$H/m \gg 1$, thus {\it violating} our basic assumption of the axion as
a coherently oscillating scalar field. Although the
numerical integration of the set of equations is possible, the equations
are not valid in this case.
In the small axion mass we are interested in, such a violation is inevitable
at some epoch in the early universe (Sec.\ \ref{sec:violation}).
To avoid this difficulty, as a simple solution, though {\it ad hoc},
we can introduce an evolving axion mass model
(Sec.\ \ref{sec:evolving}).

\section{Observational effects}
\label{sec:results}

In this section, we present our numerical calculations of background
and perturbation evolutions in the axion dark matter model with
extremely low mass.
As a fiducial model, we use the flat $\Lambda\textrm{CDM}$ model consistent
with the Wilkinson Microwave Anisotropy Probe
7-year observation
($\Omega_{b0}h^2=0.02260$, $\Omega_{c0}h^2=0.1123$, $\Omega_\Lambda=0.728$,
$n_s=0.963$, $\tau=0.087$) including massless neutrinos
with $N_\nu=3.04$ (see Table 14 of Ref.\ \cite{Komatsu-etal-2011}).
As the dark matter component we consider the axion
with $\Omega_{a0}=\Omega_{c0}$.
In order to calculate the matter and CMB power spectra, we solve
a system composed of radiation, baryon, and axion in the axion-comoving gauge.
The radiation components are handled by the Boltzmann equations
while the photon-baryon fluid by the tight coupling approximation.
The detailed numerical methods are presented in Ref.\ \cite{Hwang-2001}.

\subsection{Violation of axion-fluid nature in the early era}
\label{sec:violation}

\begin{figure}
\mbox{\epsfig{file=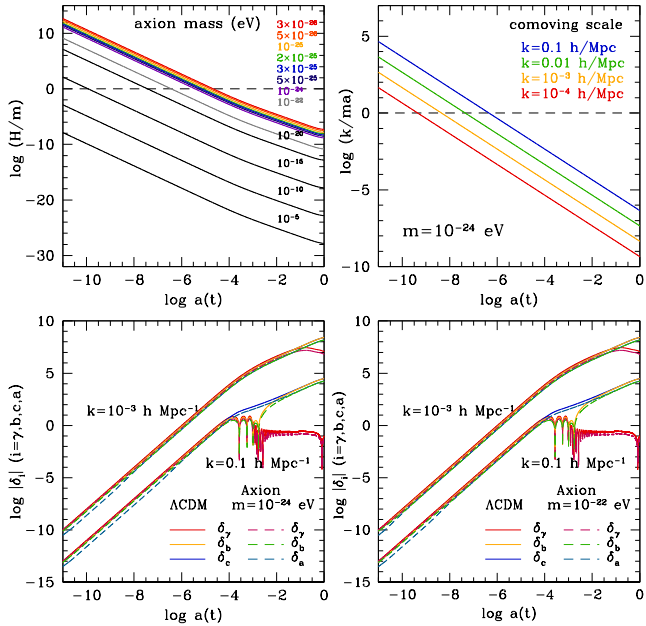,width=85mm,clip=}}
\caption{Top panels: evolution of $H/m$ (left) and  $k/(ma)$
         (right panel) as a function of scale factor $a(t)$
         for different choices of axion mass $m$ and comoving wave number $k$.
         The left panel shows that for $m = 10^{-24}~\textrm{eV}$,
         our basic assumption of $H/m \ll 1$ is {\it violated} at
         $a \lesssim 10^{-5.5}$. On the right panel showing the
         evolution of $k/(ma)$ for axion mass $m=10^{-24}~\textrm{eV}$,
         four different comoving scales from $10^{-4}$ to
         $0.1~h\textrm{Mpc}^{-1}$ have been chosen.
         Bottom panels: typical evolution of density perturbations $\delta_i$
         ($i=a$, $b$, $c$, $\gamma$) in the axion-comoving gauge.
         We considered the $\Lambda\textrm{CDM}$ model (solid curves)
         and the $\Lambda$-axion-DM model with different choices of axion mass
         (dashed curves).
         Here we consider two cases of axion mass, $m=10^{-24}~\textrm{eV}$
         (left) and $10^{-22}~\textrm{eV}$ (right panel).
         In both cases, compared with the cold dark matter in the
         $\Lambda\textrm{CDM}$ model ($\delta_c$; blue solid curve),
         the axion density perturbation ($\delta_a$; blue dashed curve)
         starts to evolve with lower amplitude; see Eq.\ (\ref{eq:ic2}).
         The behavior of $\delta_a$ slightly changes at a certain transition
         epoch
         which corresponds to when $k/(ma)$ becomes unity.
         }
\label{fig:Hm_kma_and_delta_imassv0}
\end{figure}


Due to the time dependence of $H$, our basic assumption of $H/m \ll 1$
for axion as a coherently oscillating scalar field is {\it violated}
in the early era especially for a low-mass axion.
Top panels of Fig.\ \ref{fig:Hm_kma_and_delta_imassv0}
shows evolutions of $H/m$ and $k/(ma)$ for different values of axion mass
and comoving scales.
Despite such a violation of the basic axion-fluid assumption,
here we present the baryon matter density and CMB anisotropy
power spectra for axion with a low mass.  In section \ref{sec:evolving}
we will introduce models which avoid such a violation,
and will show that the results are similar.

Examples of evolution of individual density perturbation variables are
shown in the bottom panels of Fig.\ \ref{fig:Hm_kma_and_delta_imassv0}
(see the caption for detailed descriptions), where the initial conditions for
$k/(ma) \gg 1$ limit have been used for the axion mass and the comoving scales
considered here.
The violation of $H/m \ll 1$ together with the large-scale assumption leads to
the transition between $k/(ma) \ll 1$ and $\gg 1$ at the initial epoch
(see Fig.\ \ref{fig:Hm_kma_and_delta_imassv0}, top-right panel).
In our calculation we have imposed initial conditions for $k/(ma) \gg 1$
presented in Eq.\ (\ref{eq:ic2}) because in most
cases of comoving scale and axion mass considered the values of $k/(ma)$
are larger than unity.
However, we have checked that the differences in power spectra
obtained with different sets of initial conditions in Eqs.\ (\ref{eq:ic1})
and (\ref{eq:ic2}) are negligibly small.
We also note that if we apply the initial conditions depending on the value of
$k/(ma)$ it induces a significant discontinuity in the final power spectra.

\begin{figure}
\mbox{\epsfig{file=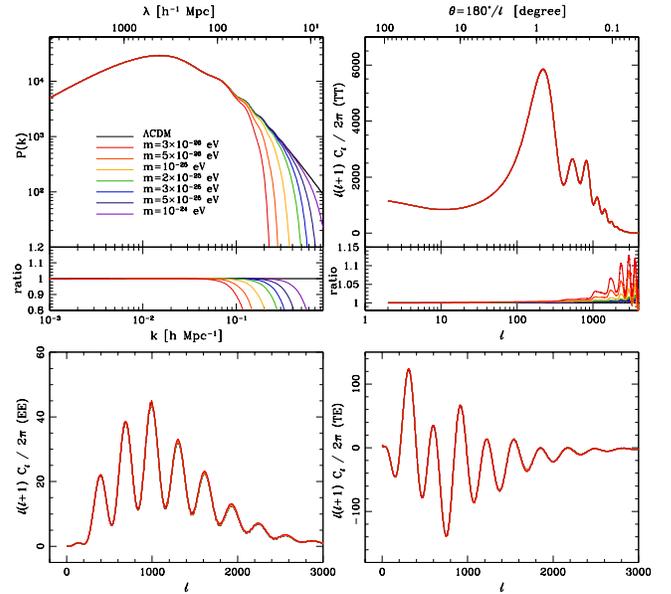,width=85mm,clip=}}
\caption{Baryonic matter (top-left) and CMB temperature, $E$-mode polarization,
         $TE$-cross power spectra (other panels)
         for axion mass $3\times 10^{-26}~\textrm{eV} \le m
         \le 10^{-24}~\textrm{eV}$ (colored curves).
         On the top-panels, the ratio of the power spectrum amplitude
         relative to the fiducial $\Lambda\textrm{CDM}$ prediction
         (black curves) are shown to indicate the differences.
         As the mass decreases below $10^{-24}~\textrm{eV}$
         we have strong small-scale suppression of the matter power spectrum
         due to the increase of axion Jeans scale.
         The CMB temperature power spectrum show small changes in higher
         multipoles corresponding to the scales below the axion Jeans scale.
         Compared with the baryon power spectrum, the change in the CMB power
         spectrum is minor.
         No noticeable changes appear in the polarization power spectra.
         }
\label{fig:matter_CMB_PS}
\end{figure}

The matter and CMB power spectra expected in the low-mass axion model
with a mass range $3\times 10^{-26}~\textrm{eV} \le m \le 10^{-24}~\textrm{eV}$
are shown in Fig.\ \ref{fig:matter_CMB_PS}.
As expected the small axion mass induces a significant damping
in the baryon matter density power spectrum.
As the axion mass decreases, the axion Jeans scale increases.
From Eq.\ (\ref{eq:Jeans_scale2}) we can see that for
$m \sim 10^{-25}~\textrm{eV}$ we have the current Jeans scale
$\lambda_J \sim 2.4~\textrm{Mpc}$.
However, the damping of the matter power spectrum appears
at the comoving scale larger than the current Jeans scale.
Note that the critical scale where the damping occurs
is determined by the Jeans scale at the radiation-matter equality.
Since the comoving Jeans wave number scales with the scale factor as
$k_{J}\propto a^{1/4}$ during the matter era, we expect
$k_{J\textrm{eq}} \approx 0.15~h\textrm{Mpc}^{-1}$
($\lambda_{J\textrm{eq}} \approx 40~h^{-1}\textrm{Mpc}$) for
$\Omega_{a0}=0.27$ and $a_{\textrm{eq}} \simeq 3\times 10^{-4}$
(see Fig.\ \ref{fig:matter_CMB_PS}), see also \cite{Hu-2000}.
Similarly, the CMB power spectra shows the deviation at
scales smaller than the axion Jeans scale, thus at very high multipoles;
for $m=3\times 10^{-26}~\textrm{eV}$, the damping of matter
power spectrum occurs around at $k_d \gtrsim 0.1~h\textrm{Mpc}^{-1}$,
which translates into the angular scales
$\ell=r_{\textrm{dec}} k_d \gtrsim 1000$
(with $r_{\textrm{dec}} \approx 10^{4}~h^{-1}\textrm{Mpc}$ the distance
to the decoupling surface); the deviation increases up to
15\% at $\ell \sim 3000$ compared to the $\Lambda\textrm{CDM}$ model
(see Refs.\ \cite{Rodriguez-Montoya-2010,Marsh-2011}
for predictions of the CMB anisotropy power spectra based on the BEC and
the ultra-light scalar field dark matter models).

\subsection{Evolving axion mass model}
\label{sec:evolving}

\begin{figure}
\mbox{\epsfig{file=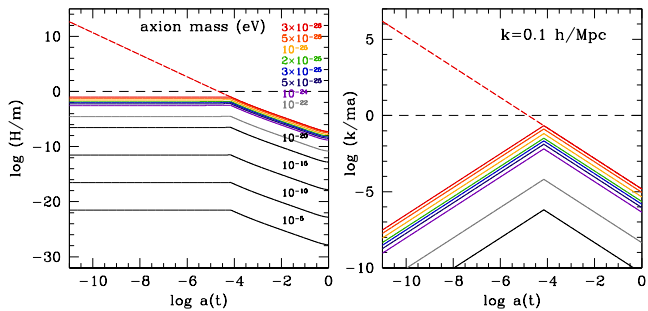,width=85mm,clip=}}
\caption{
         Evolution of $H/m$ (left) and $k/(ma)$ (right panels)
         as a function of scale factor $a(t)$ for evolving axion mass model
         [Eq.\ (\ref{eq:evolving_mass})].
         Here $a_\textrm{tr}=7\times 10^{-5}$ has been adopted and
         the axion mass ranges over $3\times 10^{-26}~\textrm{eV} \le m_0 \le
         10^{-5}~\textrm{eV}$. 
         On the right panel the variation of
         $k/(ma)$ is shown for a fixed comoving scale
         $k=0.1~h\textrm{Mpc}^{-1}$
         and for axion mass $3\times 10^{-26} \le m_0 \le 10^{-20}~\textrm{eV}$.
         For comparison, the case of constant axion mass with
         $m=3\times 10^{-26}~\textrm{eV}$ is indicated by red dashed curves.
         }
\label{fig:Hm_kma_imassv3}
\end{figure}

We have observed that the condition for axion as a rapidly oscillating scalar
field is violated in the early universe.
As a simple solution to avoid this problem, we can introduce the evolving
axion mass in the early universe.
In this way we can investigate the role of constant-low-mass axion in the
later evolution of cosmic structures within the axion-fluid formalism.
One simple mass-evolution model is to assume that the axion mass varies
in proportional to the Hubble
parameter before a transition epoch $a_\textrm{tr}$ and becomes
constant after the transition.
In the early epoch of radiation dominated era, the Hubble parameter
varies with $a^{-2}$. Thus, assuming the transition occurs in the radiation era,
the evolving axion mass model can be written as
\begin{equation}
m(a)= \left\{
   \begin{array}{cl}
         m_0 (a_\textrm{tr}/a)^2 & a<a_\textrm{tr},  \\
         m_0                     & a \ge a_\textrm{tr},
   \end{array} \right.
\label{eq:evolving_mass}
\end{equation}
where $m_0$ is the constant axion mass after the transition $a_\textrm{tr}$.
Alternatively, we may introduce a evolving axion mass model with smooth
variation with e.g., $m(a)=m_0 [1+(a_\textrm{tr}/a)^2]$.
Although not shown here, the results are quite similar to those from the simple
evolving mass model adopted here.
For the results shown in this section, we adopt
$a_\textrm{tr}=7\times 10^{-5}$, which has been set from the condition
that $H/m \approx 0.1$ at $a< a_\textrm{tr}$ for axion mass
$m_0=3\times 10^{-26}~\textrm{eV}$, the lowest axion mass considered
in this work.
Figure \ref{fig:Hm_kma_imassv3} shows evolution of $H/m$ and $k/(ma)$
for different axion mass that varies with this functional form.

\subsection{Effects of small-scale damping on power spectra and growth rate}

\begin{figure*}
\mbox{\epsfig{file=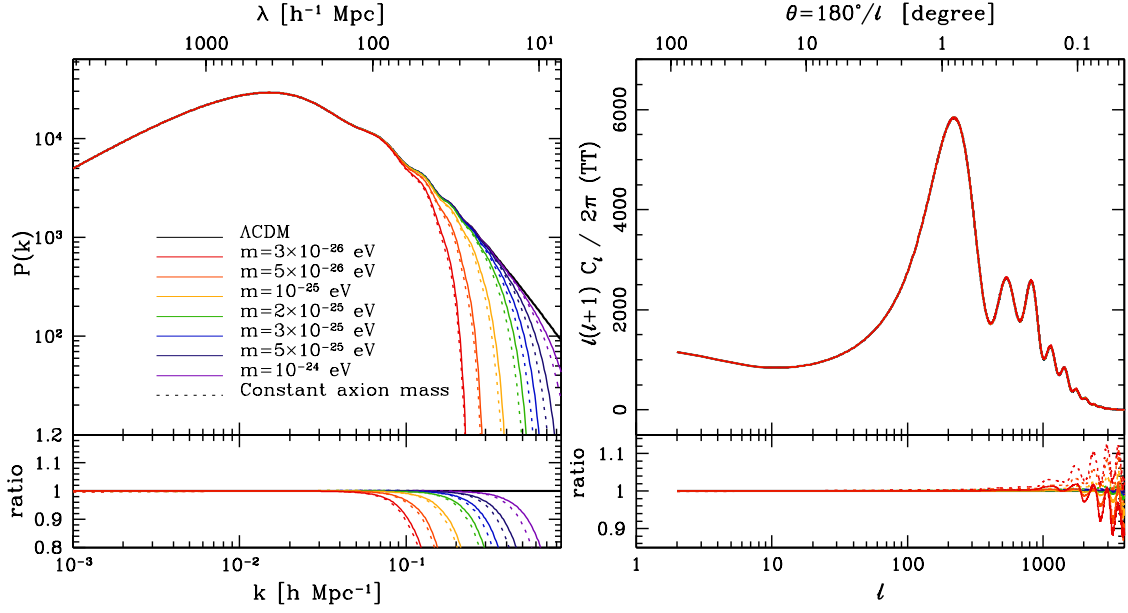,width=150mm,clip=}}
\caption{Baryonic matter (left) and CMB temperature anisotropy (right) power
         spectra for evolving axion mass model
         with $a_\textrm{tr}=7\times 10^{-5}$ for different axion masses
         (solid curves). Dotted curves represent the similar power spectra
         obtained when the constant axion mass is assumed which are
         the same as those in top panels of Fig.\ \ref{fig:matter_CMB_PS}.
         The bottom panels show power ratios relative to
         the fiducial $\Lambda\textrm{CDM}$ model prediction.
         In the CMB power spectrum the deviations at high
         multipoles ($\ell \gtrsim 1000$) show
         opposite trends depending on the constant mass and the evolving mass models.
         }
\label{fig:ps_mvar_imassv3_v2}
\end{figure*}

Here we present the baryonic matter and the CMB anisotropy
power spectra, evolution of perturbation variables, and the growth factor
based on the simple evolving axion mass model in Eq.\ (\ref{eq:evolving_mass}). 
The results for the power
spectra are shown in Fig.\ \ref{fig:ps_mvar_imassv3_v2} (solid curves),
where the power spectra for the constant axion mass (dotted curves;
the same as in Fig.\ \ref{fig:matter_CMB_PS}) are shown for comparison.

Compared to the cases of constant axion mass, the power spectra for
varying axion mass shows a small difference: the damping of the power
spectrum is slightly less significant as expected.
The damping in the matter power spectrum depends on the transition
epoch $a_\textrm{tr}$; more recent transition gives smaller damping.
Compared with the $\Lambda\textrm{CDM}$ fiducial model, the deviation
in the CMB anisotropy power spectrum at high multipole $\ell$ is now
to the negative direction. Note that in the case of constant axion mass
the deviation appears in the positive direction. 
In both cases, the magnitude of the maximum deviations
are very similar to each other.

\begin{figure}
\mbox{\epsfig{file=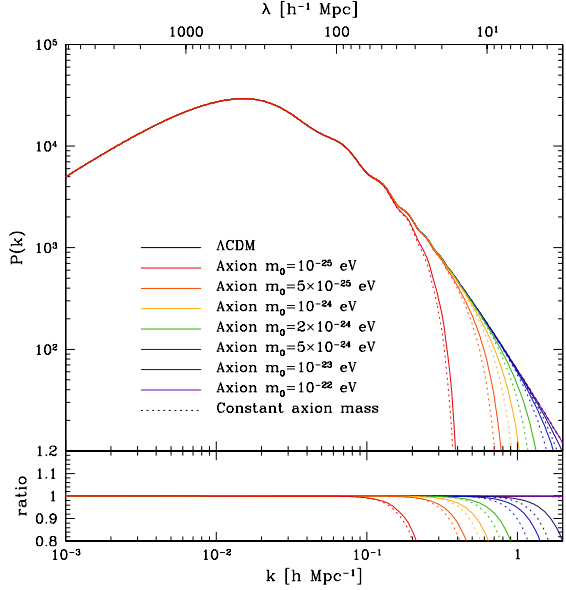,width=75mm,clip=}}
\caption{Baryonic matter density power spectra expected in the evolving
         axion mass model with 
         $10^{-25}~\textrm{eV} \le m_0 \le 10^{-22}~\textrm{eV}$ 
         (solid curves). 
         Now the range of comoving wave number has been extended to the
         higher values (smaller scales) around $k=2~h\textrm{Mpc}^{-1}$.
         Similarly to Fig.\ \ref{fig:ps_mvar_imassv3_v2}, the power spectra 
         for the constant axion mass are shown as dotted curves.
         The bottom panel shows the ratio of power relative to the
         $\Lambda\textrm{CDM}$ prediction, indicating the power spectrum
         damping on small scales.
         }
\label{fig:ps_highk}
\end{figure}

In order to determine the low-mass limit of axion we present the baryon
density power spectrum in Fig.\ \ref{fig:ps_highk}, which shows the substantial
small-scale cut-off in the power spectrum for $m < 10^{-22}~\textrm{eV}$.
For $m \ge 10^{-22}~\textrm{eV}$ the matter power spectrum is almost identical
to the CDM case on the scales considered
($\lambda \lesssim 3~h^{-1}\textrm{Mpc}$).

\begin{figure}
\mbox{\epsfig{file=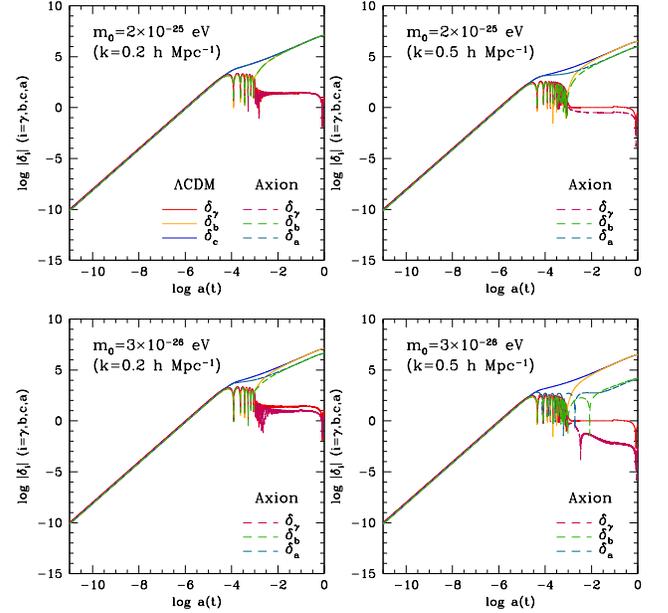,width=85mm,clip=}}
\caption{Evolution of density perturbation variables for photon, baryon,
         dark matter for evolving axion mass model, Eq.\
         (\ref{eq:evolving_mass}) with $a_\textrm{tr}=7\times 10^{-5}$.
         Shown are typical evolutions of density perturbations $\delta_i$
         ($i=a$, $b$, $c$, $\gamma$) in the axion-comoving gauge
         for $m_0=2\times 10^{-25}$, $3\times 10^{-26}~\textrm{eV}$
         and $k=0.2$, $0.5~h\textrm{Mpc}^{-1}$ (dashed curves).
         We also considered the $\Lambda\textrm{CDM}$ model whose density
         perturbations are shown as solid curves in each panel.
         Compare the results with those in Fig.\
         \ref{fig:Hm_kma_and_delta_imassv0} (bottom panels)
         }
\label{fig:delta_i_imassv3_v2}
\end{figure}
Figure \ref{fig:delta_i_imassv3_v2} shows the evolutions of density
perturbation variables for photon, baryon, and dark matter in the evolving
axion mass model at small comoving scales ($k=0.2$ and $0.5~h\textrm{Mpc}^{-1}$)
and for two extremely small axion mass ($m_0=3\times 10^{-26}$ and
$2\times 10^{-25}~\textrm{eV}$).
In all cases, the initial conditions for $k/(ma) \ll 1$ limit have been used.

\begin{figure}
\mbox{\epsfig{file=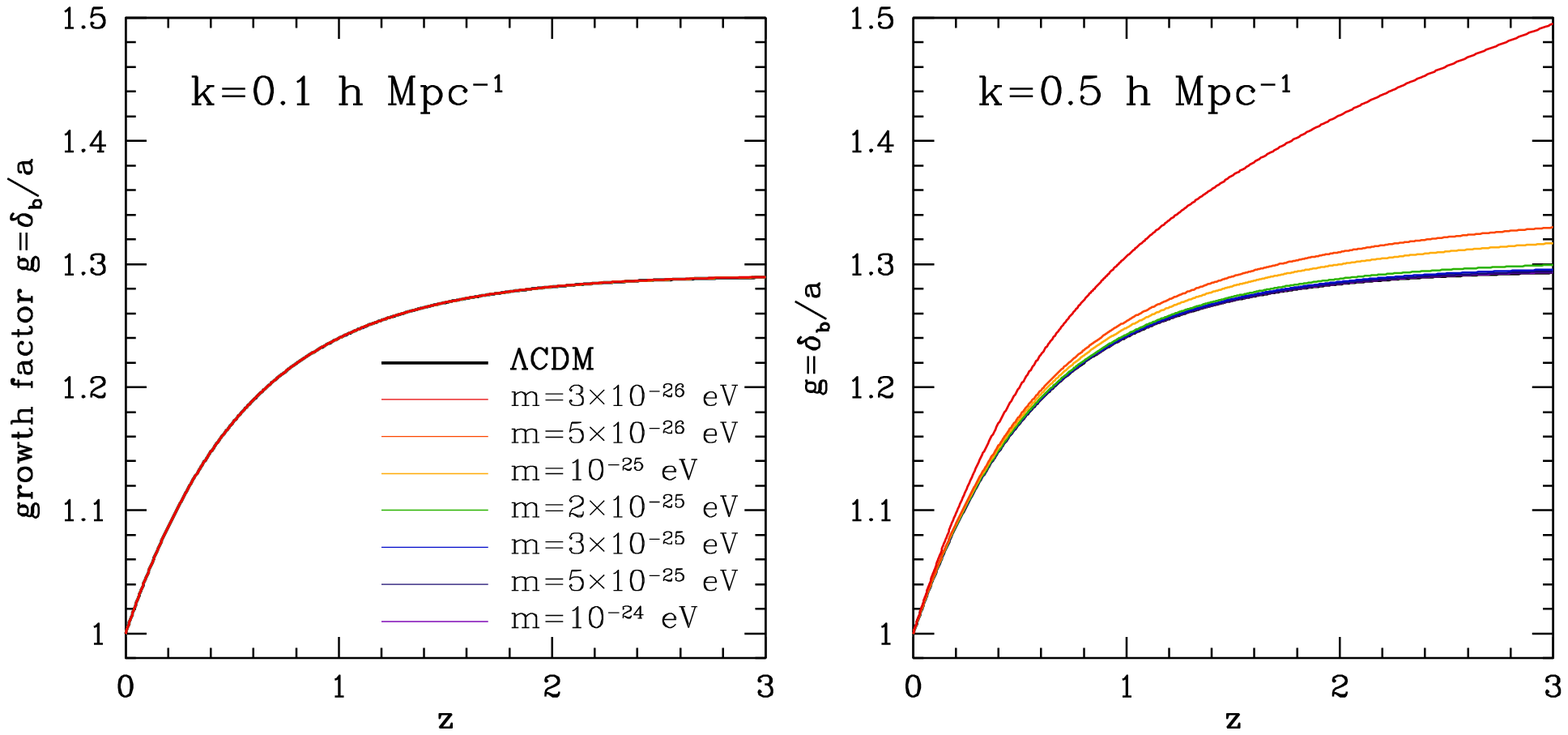,width=85mm,clip=}}
\caption{Evolution of the normalized perturbation growth factor
         $g\equiv (\delta_b /a)$ at comoving scales of $k=0.1$,
         $0.5~h\textrm{Mpc}^{-1}$ for different axion masses
         (colored curves). Here we use an evolving-axion-mass model
         with $a_\textrm{tr}=7\times 10^{-5}$.
         Black curves represent the growth factor of the fiducial
         $\Lambda\textrm{CDM}$ model.
         }
\label{fig:ggg_imassv3}
\end{figure}

In Fig.\ \ref{fig:ggg_imassv3} we show the growth factor
$g \equiv \delta_b /a$ (normalized to unity at present)
at comoving scales $k=0.1$, $0.5~h\textrm{Mpc}^{-1}$ and
for various axion mass values. At $k=0.5~h\textrm{Mpc}^{-1}$,
noticeable positive deviations from the $\Lambda\textrm{CDM}$ prediction
are seen.  
Despite the significant suppression of overall perturbation growth,
the recent perturbation growth factor is larger in the low-mass axion
model, especially at the small comoving scales.
This can be interpreted as the more growth in the axion mass model
(see Refs.\ 
\cite{Suarez-2011,Kain-2012,Chavanis-2012,Magana-2012-review}).

\subsection{Mixed cold and hot dark matter model}
\label{sec:HDM}

\begin{figure}
\mbox{\epsfig{file=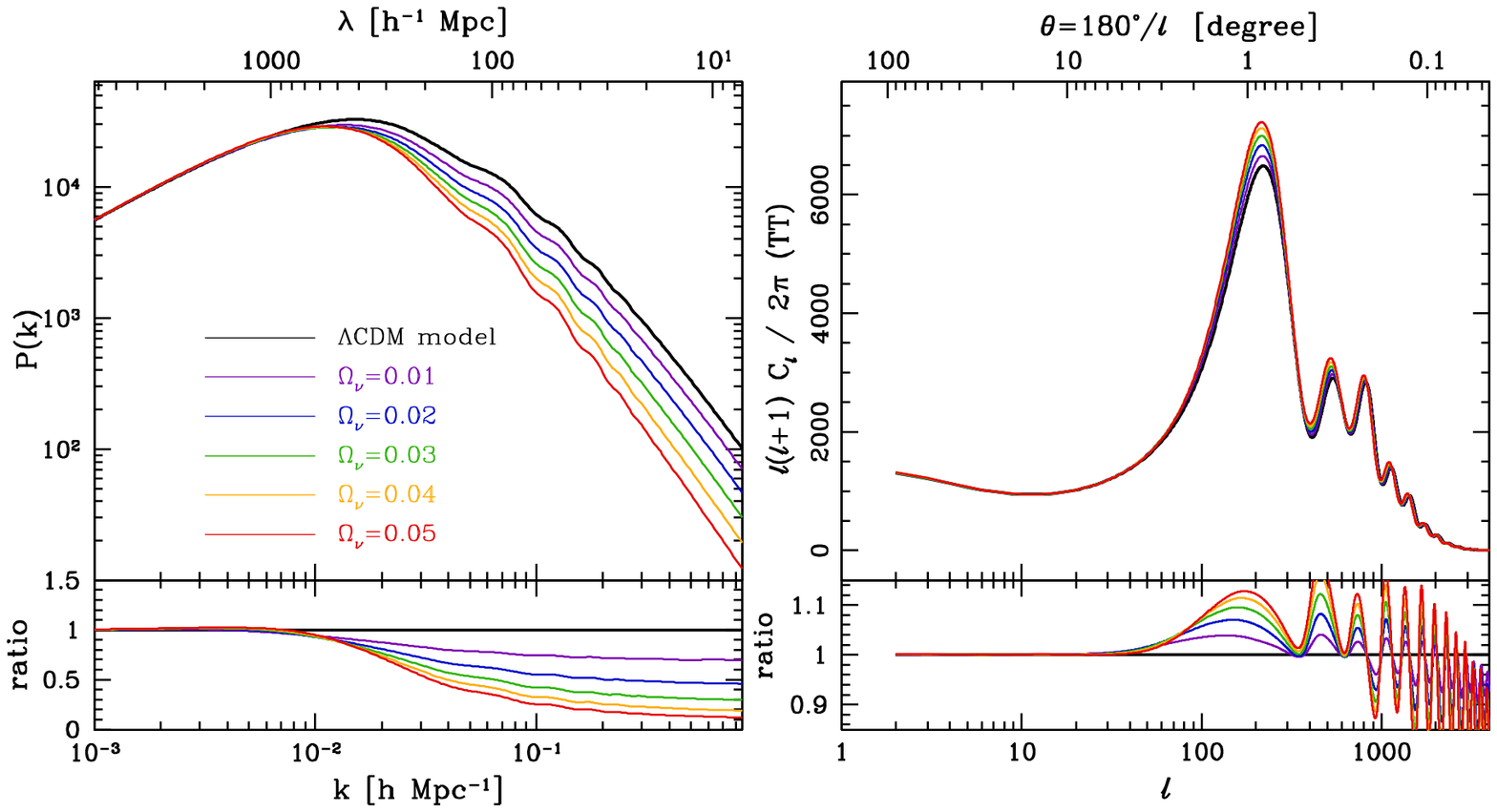,width=85mm,clip=}}
\caption{Matter and CMB temperature anisotropy power spectra
         expected in a model containing small fraction of massive neutrino
         component, where we have used the CAMB software \cite{CAMB}.
         Here we consider massive neutrinos ($N_\nu=3.04$)
         whose mass ranges from
         $m_\nu=0.154~\textrm{eV}$ ($\Omega_{\nu 0}=0.01$; red)
         to $0.769~\textrm{eV}$ ($\Omega_{\nu 0}=0.05$; violet curves),
         with a relation $(\Omega_{\nu 0}+\Omega_{c0})h^2=0.1123$.
         Black curves represent the power spectrum of the fiducial
         $\Lambda\textrm{CDM}$ model with massless neutrinos.
        The curves in the bottom panels indicate the ratios of powers
         relative to the $\Lambda\textrm{CDM}$ model prediction.
         }
\label{fig:pow_massive_neutrino}
\end{figure}

The warm dark matter model or the mixture of the cold and hot dark matter model 
are the alternative candidates to the
CDM model with small-scale suppression in the matter power spectra.
It would be interesting to compare the small-scale suppression of our
low-mass axion model with those models.
In Fig.\ \ref{fig:pow_massive_neutrino} we present the case of massive
neutrinos contributing as the hot dark matter added to the dominant CDM model.
The results are similar to those found in Ref. \cite{Lesgourgues-2006}.

Although the low-mass axion dark matter model shows a sharp damping
in the baryonic matter power spectrum, the CDM model with a small fraction
of massive neutrinos gives a damping of power with almost constant factor at
$k > 0.1~h\textrm{Mpc}^{-1}$. Besides, the CMB anisotropy power spectrum
is very sensitive to the massive neutrino contribution; the massive neutrinos
affect all the acoustic oscillatory features at intermediate and
high angular scales ($\ell \gtrsim 100$). On the other hand, the behavior of
the axion dark matter model is quite different, and is almost insensitive
to the axion mass except at the higher multipoles.
In the low-mass axion case the changes from the CDM in both power spectra 
occur only at scales smaller than the axion Jeans scale.

\section{Discussion}
\label{sec:discussion}

In this work, we have studied effects of extremely low-mass
($m \le 10^{-24}~\textrm{eV}$) axion on the baryon matter density
and the CMB anisotropy power spectra, and the perturbation
growth based on the full relativistic linear perturbation analysis.
With a low mass, however, the basic assumption about the coherently
oscillating scalar field is inevitably broken in the early universe
($H/m \gg 1$).
We have introduced the simple evolving-axion-mass-in-the-early-universe model
to avoid this problem.

We showed that axion mass smaller than $10^{-24}~\textrm{eV}$ induces the
characteristic significant damping in the baryon density power spectrum 
on scales smaller than the axion Jeans scale, and changes in the higher
multipole in the CMB anisotropy power spectra.
Except for small changes in the higher multipoles ($\ell \gtrsim 1000$)
corresponding to the scales smaller than the axion Jeans scale, the CMB
power spectrum remains the same as the CDM case. 
The CDM nature is also preserved in the baryon matter power spectrum above
the axion Jeans scale. We showed that the small-scale damping nature of
our low-mass axion model differs from the one expected in the CDM model
mixed with the massive neutrinos as a hot dark matter component.

Whether the small-scale damping of the baryon matter density
power spectrum can help alleviating the excess small-scale clustering problem
of the CDM model requires further studies in the nonlinear clustering
properties of the light mass axion.

%
%
\noindent{\bf Acknowledgments:}
J.H.\ was supported by KRF Grant funded by the Korean Government
(KRF-2008-341-C00022). H.N.\ was supported by grant No.\
2010-0000302 from KOSEF funded by the Korean Government (MEST).

\def\and{{and }}


\end{document}